\documentclass[sigconf]{acmart} 
\usepackage{tabularx}
\usepackage{booktabs} 
\usepackage{makecell} 

\AtBeginDocument{%
  }

\setcopyright{acmlicensed}
\copyrightyear{2018}
\acmYear{2018}
\acmDOI{XXXXXXX.XXXXXXX}

\acmConference[Conference acronym 'XX]{Make sure to enter the correct
  conference title from your rights confirmation emai}{June 03--05,
  2018}{Woodstock, NY}
\acmISBN{978-1-4503-XXXX-X/18/06}
\usepackage{subcaption} 
\usepackage[ruled,linesnumbered]{algorithm2e}
\usepackage{algpseudocode}

\begin{document}

\title{GANcrop: A Contrastive Defense Against Backdoor Attacks in Federated Learning}

%

\author{Xiaoyun Gan}
\affiliation{%
  \institution{Guangxi Normal University}
  \streetaddress{15 Yucai Rd}
  \city{Guilin}
  \state{Guangxi}
  \country{China}
  \postcode{541004}}
\email{ganxiaoyun@stu.gxnu.edu.cn}
\orcid{0009-0002-6154-0855}

\author{Shanyu Gan}
\affiliation{%
  \institution{Guangxi Normal University}
  \streetaddress{15 Yucai Rd}
  \city{Guilin}
  \state{Guangxi}
  \country{China}
  \postcode{541004}}
\email{gsy520@stu.gxnu.edu.cn}
\orcid{0009-0007-6014-0290}

\author{Taizhi Su}
\affiliation{%
  \institution{Guangxi Normal University}
  \streetaddress{15 Yucai Rd}
  \city{Guilin}
  \state{Guangxi}
  \country{China}
  \postcode{541004}}
\email{csstz@stu.gxnu.edu.cn}
\orcid{0009-0003-7646-5664}

\author{Peng Liu*}
\affiliation{%
  \institution{Guangxi Normal University}
  \streetaddress{15 Yucai Rd}
  \city{Guilin}
  \state{Guangxi}
  \country{China}
  \postcode{541004}}
\email{liupeng@gxnu.edu.cn}
\orcid{0000-0003-2583-9112}



%
\renewcommand{\shortauthors}{Trovato et al.}

\begin{abstract}
  With heightened awareness of data privacy protection, Federated Learning (FL) has attracted widespread attention as a privacy-preserving distributed machine learning method. However, the distributed nature of federated learning also provides opportunities for backdoor attacks, where attackers can guide the model to produce incorrect predictions without affecting the global model training process.
  This paper introduces a novel defense mechanism against backdoor attacks in federated learning, named GANcrop. This approach leverages contrastive learning to deeply explore the disparities between malicious and benign models for attack identification, followed by the utilization of Generative Adversarial Networks (GAN) to recover backdoor triggers and implement targeted mitigation strategies. Experimental findings demonstrate that GANcrop effectively safeguards against backdoor attacks, particularly in non-IID scenarios, while maintaining satisfactory model accuracy, showcasing its remarkable defensive efficacy and practical utility.
  
\end{abstract}




\maketitle

\section{Introduction}
With the rise in awareness of data privacy protection, centralized data collection faces significant challenges, making collecting training data a pressing issue in machine learning. Federated learning, as a novel distributed machine learning method that protects privacy \cite{1}, cleverly bypasses data collection challenges and has thus received widespread attention.

In Federated Learning (FL), training data is kept locally on the user's device, and only the model initialization and trained model parameters are transmitted between the server and participating users. This method allows for secure training collaboration among multiple parties while protecting user privacy. Nevertheless, owing to the data being distributed among various participating entities, this distributed characteristic, while rendering the model training more flexible and secure, concurrently offers attackers a foothold\cite{2,3}. They can stealthily guide the model to produce specific predictions for certain triggers without impacting the global model training process \cite{4}, thereby diminishing the model's accuracy and credibility. Thus, researching effective defenses against backdoor attacks in federated learning is crucial for ensuring the model's security and reliability.

In the field of deep learning, there has been considerable research dedicated to defending against backdoor attacks. Methods for recovering backdoor triggers, such as NC \cite{5} and GANsweep \cite{6}, aim to mitigate backdoor attacks by recovering the triggers of poisoned models, thereby achieving defense objectives. However, due to the specific data distribution in federated learning, applying methods from centralized learning directly to FL may lead to poor defense outcomes or model performance, often resulting in the global model failing to converge. Furthermore, the model updating process in federated learning involves parameter exchange and model aggregation among servers and multiple users, which adds to the defense's complexity. Therefore, developing new defense mechanisms tailored to the peculiarities of federated learning is necessary to address potential backdoor attack threats.

Existing federated learning backdoor defences schemes can be mainly divided into two categories: methods based on anomaly detection and those utilizing pruning or noise addition techniques. Methods based on anomaly detection usually require extensive computations on the server side \cite{7}, detecting backdoor attacks by monitoring the similarity between models or the abnormal changes in update behaviors. However, due to the complexity of federated learning models and the heterogeneity of data distribution, these methods often incur excessive computational costs and have high prerequisites for the scenarios \cite{8,9}. On the other hand, methods based on pruning or adding noise work by diluting or reducing the impact of malicious model updates on the global model through the insertion of noise into the model updates \cite{10,11}. Although this approach can enhance the robustness of the model, the added noise often leads to a decrease in model performance and accuracy, thus affecting the overall performance of the model. Therefore, it is imperative to seek a more effective method for defending against federated learning backdoor attacks to balance security and performance demands.

To overcome the limitations of existing methods, this paper proposes a federated learning backdoor attack defense method based on Contrastive Learning \cite{12} and Generative Adversarial Networks (GAN) \cite{13}, named GANcrop. This method utilizes contrastive learning to delve into the differences between malicious and benign models, achieving effective attack identification. Then, on the predicted malicious models, it uses GAN to recover the trigger of poisoned models and carries out targeted backdoor mitigation to achieve the defense purpose. The contributions of this paper are as follows:

\vspace{-0.5em}
\begin{itemize}

\item We have implemented a new federated learning backdoor attack defense method, GANcrop, based on contrastive learning and generative adversarial networks.

\item We introduced a model detection method based on contrastive learning, which can effectively distinguish between malicious and benign models under the multi-model scenario of federated learning, achieving effective attack detection. This is one of the rare model-level contrastive learning methods in existing research.

\item We applied the backdoor trigger recovery approach from deep learning to the defense work against backdoor attacks in federated learning. In non-IID scenarios, our method achieves sound defense effects while ensuring model accuracy.
\end{itemize}

\section{Preliminary Knowledge}
Backdoor attacks in neural networks aim to guide the neural network model to produce incorrect output results by making subtle, humanly imperceptible modifications to the input data. Attackers inject backdoors into the model during the training or deployment phase, causing the model to produce incorrect predictions under specific triggering conditions. In backdoor attacks, the attacker's task can be described as a multi-objective optimization problem, where the attacker needs to maintain the accuracy of the main task while achieving a high success rate of backdoor attacks on targeted attack class samples \cite{14}. The optimization objective function of a backdoor attack can be represented by Formula (1):

\begin{equation}
  \theta^* = \min_{\theta} \left( \sum_{i \in |D|} L(x_{_i} , y_{_i} ) + \sum_{i \in \left | D_p \right | } L(\psi(x_{_i}), \tau(y_{_i} )) \right)
\end{equation}
     
Where D is the test set for the main task;  \({D_p}\) is the poisoned dataset containing backdoor samples, these samples are manipulated by the transformation function \( \psi \), outputting a specific \( y \) under the backdoor task.

\begin{figure}[h]
  \centering
  \includegraphics[scale=0.35]{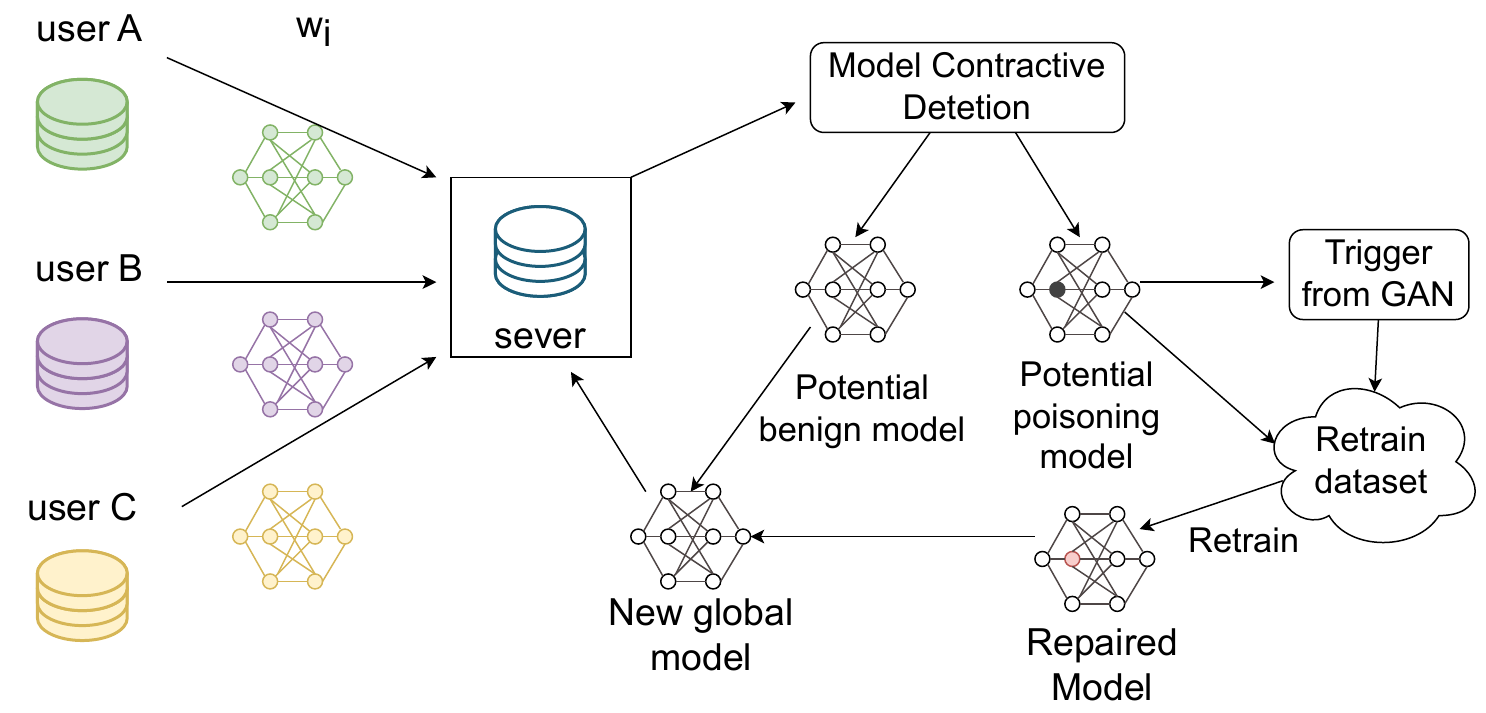}
  \caption{GANcrop architecture diagram }
  \Description{A woman and a girl in white dresses sit in an open car.}
  \label{fig1}
  \vspace{-2.0em}
\end{figure}

\section{Proposed Method}

This section will introduce the federated learning backdoor attack defense framework based on Generative Adversarial Networks (GAN). This framework mainly consists of three modules: attack detection, backdoor mitigation, and model aggregation. The framework of our method is shown in Fig. \ref{fig1}.

\subsection{Attack Detection}

In devising strategies to defend against backdoor attacks, directly comparing the similarity between models is often insufficient to identify contaminated models, as triggers are usually concealed. It is necessary to delve into the analysis of model parameter differences and sensitivities to distinguish between benign and poisoned models. In backdoor attacks, triggers are generally placed around the periphery of images, not disturbing the main task's accuracy. It's challenging to analyze the trigger location from model parameters directly, so we employ contrastive learning to train an anomaly detection model sensitive to edge position parameters, conducting direct contrastive training on model data, which is particularly crucial in federated learning.

We construct a poisoned dataset with non-central triggers to train the poisoned models. Under the contrastive learning framework, we build positive and negative sample pairs from the parameters of poisoned and clean models. Unlike traditional methods \cite{12}, in our construction, different poisoned model samples from different poisoned or benign models benign model samples are considered positive samples, while poisoned and benign models are considered negative samples. This method of constructing sample pairs allows the model to extract as many common features among poisoned models as possible. Our framework optimizes the contrastive learning model by conducting direct contrastive training on model data and learning the patterns of trigger locations to distinguish between model categories accurately. This is key to our model-level contrastive detection in federated learning.

\begin{figure}[h]
  \centering
  \includegraphics[width=1\linewidth,height=6cm]{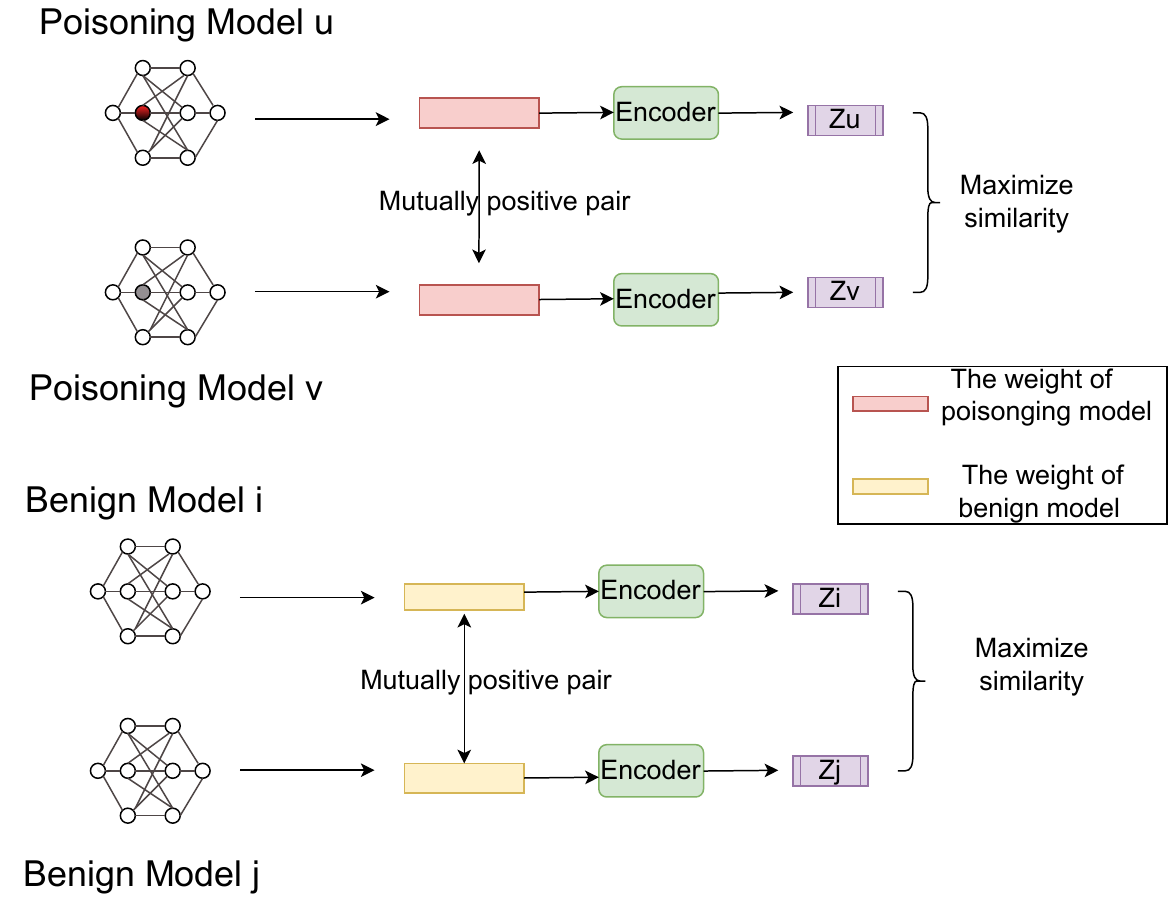}
  \caption{Model Contrastive Learning }
  \label{fig2}
\end{figure}

Our model contrast detection diagram is shown in Fig. \ref{fig2}. The specific contrastive loss function is shown in Formula (2):

\begin{equation}
L_{\text{InfoNCE}} = -\log \frac{\exp(d(u,v)^-/\tau)}{\sum_{k=1}^K \exp(d(u,v_k)^+/\tau)}
\end{equation}

Where \(d(u,v)^-\)represents the distance between positive sample pairs, \(d(u,v_k)^+\) represents the distance between negative sample pairs, and \(\tau\) is the temperature parameter of the contrastive loss, used to control the degree of softening.

Through the extraction of model representations using contrastive learning, it's possible to identify the common characteristics of potentially anomalous models. Utilizing these characteristics for model classification training allows for precisely excluding malicious influences and identifying potentially poisoned models. This training process is completed on the server side using a public dataset to distinguish potential poisoned models through feature extraction and classifier training. The algorithm for GANcrop, as shown in Algorithm \ref{alg1}.

\subsection{Backdoor Mitigation}

To effectively defend against backdoor attacks, we further mitigate backdoors based on the results of parameter comparison detection, reducing the impact of the attacks. We use GAN to recover the backdoor triggers of poisoned models, reintroduce the recovered backdoors into the server's clean dataset with correct labels and retrain the poisoned models to mitigate the backdoors.

The GANGSWEEP framework inspires the backdoor recovery strategy, a defense method in deep neural networks that uses GANs to recover backdoor triggers and mitigate their effects. However, directly applying it to FL may cause convergence difficulties for the global model. Therefore, this paper adopts a method of attack detection followed by backdoor mitigation to defend against backdoor attacks in FL.

\begin{figure}[h]
  \centering
  \includegraphics[width=1\linewidth]{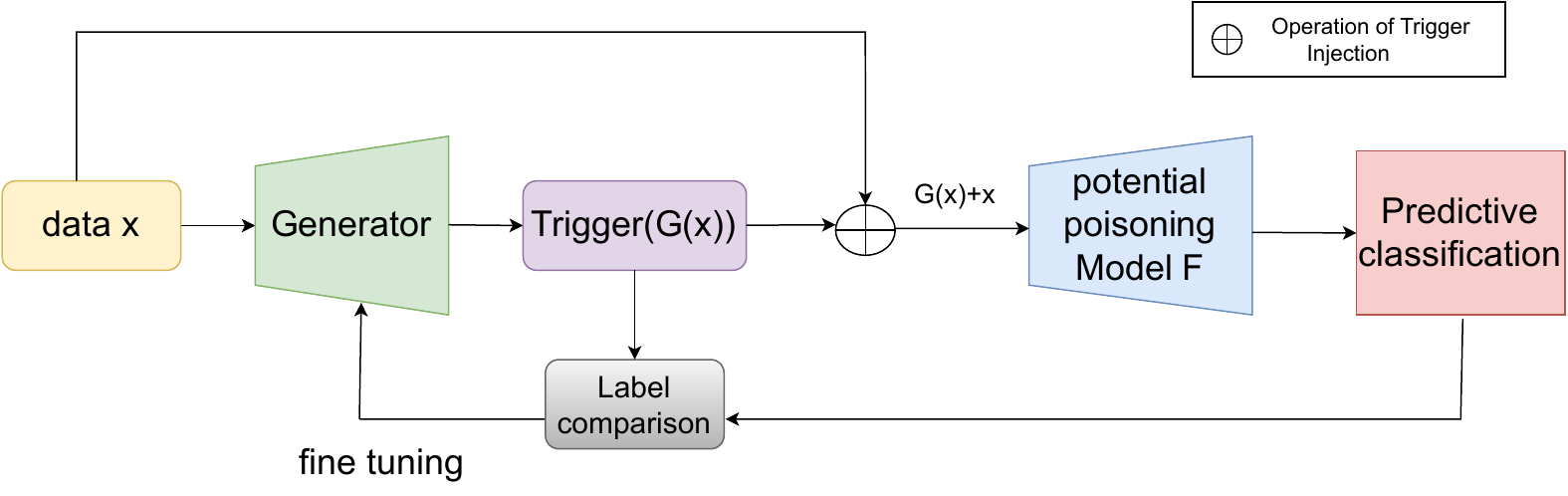}
  \caption{Trigger Generator}
  \label{fig3}
  \vspace{-2.0em}
\end{figure}

In this paper, we attempt to recover the backdoor triggers in the contaminated model F using the generator of a Generative Adversarial Network (GAN). The backdoor triggers are usually designed as special vectors with the same dimensions as the image. Such a design ensures the feasibility of generating triggers from the image model. Unlike traditional GANs, we do not optimize the discriminator during training; instead, we use the model from potentially malicious users as a substitute to guide the generator in recovering the original triggers more accurately. To verify the effectiveness of the generated triggers, this paper injects the generated triggers one by one into the images (marked as $x$) in the validation dataset, forming new images $G(x)+x$, and observes the predictions of the malicious model F on these new images to guide the generator in iteratively improving, gradually learning trigger features. Our architecture can effectively recover backdoor triggers through continuous iterative training, as illustrated in Fig. \ref{fig3}.

\begin{algorithm}[tb]
    \caption{Model Contrastive Detection Algorithm}
    \label{alg1}
    \SetAlgoLined
    \KwIn{Number of simulated clients $N$, learning rate $\eta$, ResNet network model $R$, initial feature extraction $SimModel$, simulated user model parameters $w_{i}^{t}$ , epoch $E$, trigger $TR$, public dataset $D_c$, proportion of poisoned dataset $\gamma$, simulated attack user set $S_p$, simulated benign user set $S_b$, positive and negative sample pair construction function $g(\cdot)$}
    \KwOut{Model feature extractor $SimModel$, discriminator}
    \SetKwFunction{FModContract}{ModContract}
    \SetKwFunction{FModClassify}{ModClassify}
    
    \SetKwProg{Fn}{Function}{:}{}
    \Fn{\FModContract{$D_c, TR$}}{
        $D_p = \gamma D_c + TR$\;
        Send the initial model to each simulation client\;
        \For{$t=0$ \KwTo $T-1$}{
            \For{$j = 0$ \KwTo $N-1$}{
                Send global model parameters to the customer\;
                \If{$j \in S_p$}{
                    $w_{i}^{t} \leftarrow \text{Local Training}(j,w^{t}, D_p)$\;
                }\Else{
                    $w_{i}^{t} \leftarrow \text{Local Training}(j,w^{t},(D_c-D_p))$\;
                }
                $(u, v) = g(w_{i}^{t}(j \in S_p), w_{i}^{t}(i \in S_b))$\;
                $L_{\text{InfoNCE}} = -\log \frac{\exp(d(u,v)^-/\tau)}{\sum_{k=1}^K \exp(d(u,v_k)^+/\tau)}$\;
                $SimModel \leftarrow SimModel - \eta \nabla L$\;
            }
        }
        \KwRet{IdentifyMod}\;
    }
    \SetAlgoRefName{\FModContract}
    
    \Fn{\FModClassify{$SimModel, S_p, S_b,w_{i}^{t}$}}{
        $IdentifyMod \gets$ \text{LinearEvaluate}($SimModel, \text{classes}$)\;
        \For{$j = 1$ \KwTo $E$}{
            $IdentifyMod_{j} \gets IdentifyMod_{j-1} -\eta \nabla F_j(S_p, S_b)$\;
        }
        \KwRet{IdentifyMod}\;
    }
    \SetAlgoRefName{\FModClassify}
\end{algorithm}

After obtaining the generated triggers highly similar to the original poisoned model's triggers, this paper cleverly injects these generated triggers into a new, clean dataset while keeping the dataset's real labels unchanged for model retraining. The purpose is to allow the model to correctly identify and handle the correspondence between image data and labels during retraining, thereby effectively eliminating the impact of the backdoor triggers. The retraining process is a process of knowledge updating and forgetting, where the model gradually adjusts its internal parameters, forgetting the original backdoor features to adapt to the new data distribution.

\subsection{Model Aggregation}

Based on the results of attack detection, this paper identifies and filters out benign models and potentially malicious user models. Then, through the backdoor recovery of the GAN model, pseudo triggers similar to the original triggers of potentially malicious models are generated, and these pseudo triggers are used for targeted backdoor mitigation. The models retrained afterward will be restored to a state unaffected by any malicious influence.

To enhance the robustness and reliability of the model, fully utilizing the training success of each user model, this repaired model is weighted and aggregated with the identified benign models to obtain a global, more robust federated learning model.

\section{Experiment}
\subsection{Experimental Setup}

This experiment uses the RestNet18 neural network model as the base architecture. The experiment employs a 3$\times$3 convolution kernel, with stride and padding values set to 1. This experiment omits pooling layers to avoid excessive compression of image information. The experiment involves 40 clients, selects CrossEntropyLoss as the loss function, and adopts classic stochastic gradient descent (SGD) as the optimizer, with a learning rate set to 0.1. In each iteration round, each client conducts training for four epochs.

\paragraph{Dataset:} The experiment utilizes the widely used Cifar10 dataset in the same domain applications. The dataset contains 60,000 data samples, 50,000 training samples, and 10,000 test samples, with 10 categories in the dataset.

\begin{figure}[ht]
  \centering
  \begin{subfigure}[b]{0.21\linewidth}
    \includegraphics[width=\linewidth]{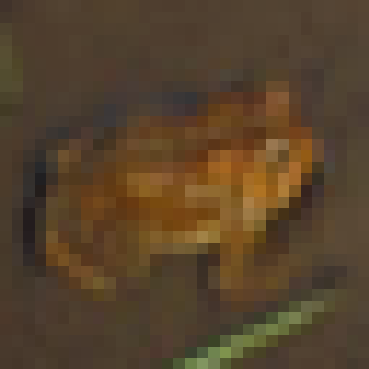}
    \caption{} 
    \label{fig:sub1}
  \end{subfigure}
  \begin{subfigure}[b]{0.21\linewidth}
    \includegraphics[width=\linewidth]{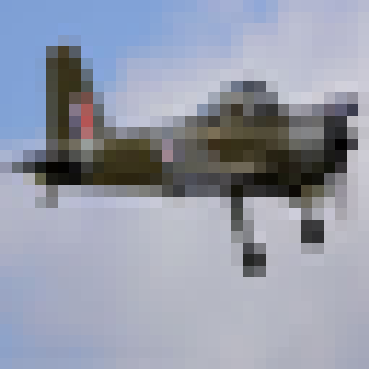}
    \caption{}
    \label{fig:sub2}
  \end{subfigure}
  \begin{subfigure}[b]{0.21\linewidth}
    \includegraphics[width=\linewidth]{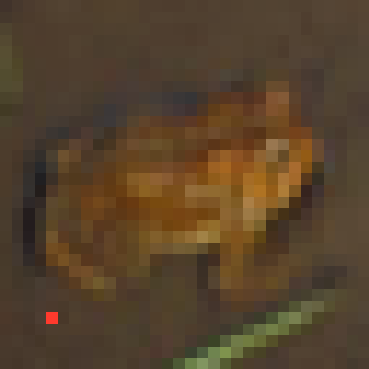}
    \caption{}
    \label{fig:sub3}
  \end{subfigure}
  \begin{subfigure}[b]{0.21\linewidth}
    \includegraphics[width=\linewidth]{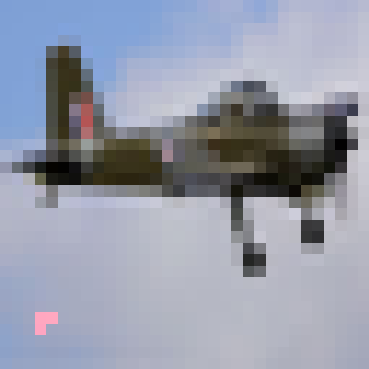}
    \caption{}
    \label{fig:sub4}
  \end{subfigure}
  \caption{Attack Schematic}
  \label{fig4}
\end{figure}

\paragraph{Data Division:} To achieve a non-IID data distribution, the experiment adopts the data skew and label skew methods proposed in the literature \cite{15}, setting the Dirichlet coefficient to 0.7 to control the degree of data skew.

\paragraph{Attack Method:} The trigger injection \cite{19} backdoor attack method is adopted. Fig. \ref{fig4} shows the effect of trigger injection using the cifar10 dataset, where the first two subfigures are original pictures, and the last two subfigures are poisoned pictures after trigger injection.

\subsection{Experimental Results and Analysis}

\begin{figure}[ht]
  \centering
  \begin{subfigure}[b]{0.45\linewidth}
     \includegraphics[width=\linewidth]{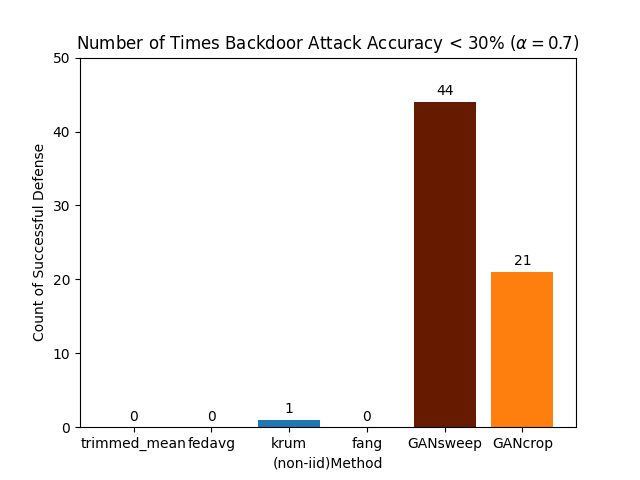}   
    \caption{} 
    \label{fig:sub1}
  \end{subfigure}
  \begin{subfigure}[b]{0.45\linewidth}
    \includegraphics[width=\linewidth]{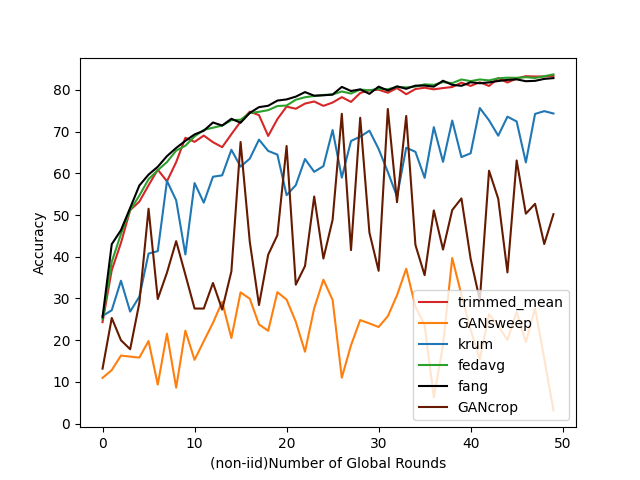}    
    \caption{}
    \label{fig:sub2}
  \end{subfigure}
  \caption{Successful rounds of defending against attacks and main task accuracy of six methods}
  \label{fig5}
  \vspace{-2.0em}
\end{figure}

\paragraph{Verifying the Effectiveness of GANcrop in Defending Against Backdoor Attacks:} To compare the performance of GANcrop, control methods were selected including FedAvg \cite{1} without defense measures as a baseline and four representative defense methods: Krum\cite{16}, Trimmed-mean \cite{17}, Fang\cite{18}, and GANsweep\cite{6}. An attack ratio of 30\(\%\) was set for the experiments, with these attacking users launching attacks in each round of iteration. 
Fig. ~\ref{fig5}~(a) depicts the rounds of successful defense against backdoor attacks for six methodologies on the CIFAR-10 dataset, across a total of 50 experimental rounds. Here, we define a successful defense against backdoor attacks as instances where backdoor accuracy falls below 30\%. Typically, the level of backdoor accuracy is directly correlated with the success rate of the backdoor attack.

The experimental results indicate that the models trained using FedAvg, Krum, Trimmed-mean, and Fang methods exhibit minimal effectiveness in defending against backdoor attacks. In contrast, GANsweep demonstrates a significantly higher defense capability when confronted with corresponding trigger attacks. On the other hand, the GANcrop scheme presented in this paper achieves a more balanced effect in defending against backdoor attacks. The reason is that GANcrop is a federated scheme aiming to balance maintaining the accuracy of the model's main task with reducing the success rate of attacks.

\paragraph{Verifying the main task accuracy of GANcrop:} Fig. ~\ref{fig5}~(b) shows the change in the main task accuracy of the six experimental methods on the cifar10 dataset. FedAvg, Trimmed-mean, and Fang do not effectively defend against backdoor attacks despite having high main task accuracy. However, combined with the experimental results of backdoor accuracy above, it's clear that FedAvg, Trimmed-mean, and Fang, despite having high main task accuracy, do not effectively defend against backdoor attacks. By discarding a portion of user models during the global aggregation phase, Krum shows weakness in main task accuracy. Meanwhile, GANsweep, although it has backdoor solid defense capabilities, also leads to lower main task accuracy.

Additionally, the GANcrop method presents a compromise in terms of the model's main task accuracy, achieving a level of backdoor defense effectiveness while maintaining a main task accuracy that is higher than that of the GANsweep method but lower than the other four methods.

\begin{figure}[ht]
  \centering
  \begin{subfigure}[b]{0.45\linewidth}
    \includegraphics[width=\linewidth]{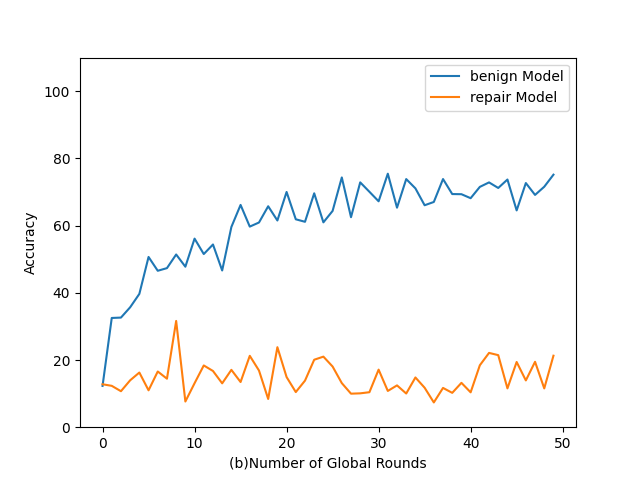}
    \caption{} 
    \label{fig:sub1}
  \end{subfigure}
  \begin{subfigure}[b]{0.45\linewidth}
    \includegraphics[width=\linewidth]{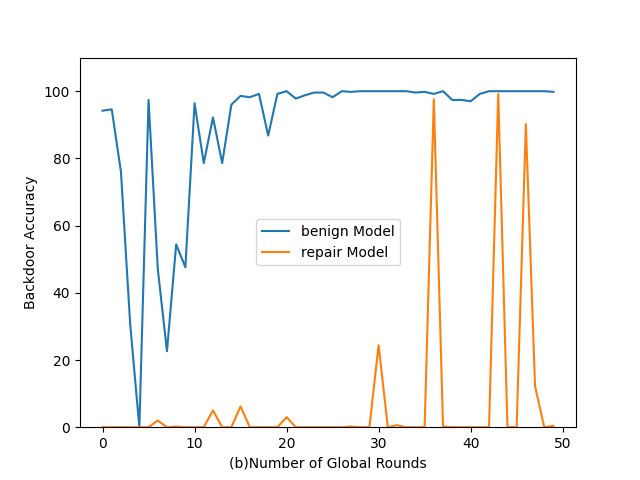}
    \caption{}
    \label{fig:sub2}
  \end{subfigure}
  \caption{The main task accuracy and backdoor attack success rate of two submodels}
  \label{fig6}
  \vspace{-2em}
\end{figure}

\paragraph{Verify the main task accuracy and the backdoor task accuracy of the two sub-models:}
To validate the model's attack detection and repair performance, this paper analyzed the main task accuracy and the backdoor task accuracy of two sub-models. Fig. ~\ref{fig6}  shows these results: Fig.~\ref{fig6}~(a)  displays the main task accuracy of both the benign prediction model and the repair model, while Fig. ~\ref{fig6}~(b) presents their backdoor task accuracy. Among them, a higher backdoor task accuracy represents a higher backdoor attack success rate.  Fig.~\ref{fig6}~(a) reveals that the benign model outperforms the repair model in main task accuracy, potentially due to the repair model's dataset. Fig. ~\ref{fig6}~(b) shows that the backdoor task accuracy is significantly lower in the repair model. This highlights the importance of integrating the benign and repair models to lessen the main task accuracy loss from backdoor mitigation and diminish the effects of undetected attacks.

\paragraph{Verifying Execution Time of Six Approaches:} The experiment further compared the execution time of the six different methods, aiming to assess their total time consumption from local computation to global aggregation in a single round of iteration. The execution times obtained through experimental verification are shown in Table \ref{table1}.

\begin{table}    
  \caption{Execution Time}    
  \label{table1}    
  \small 
  \begin{tabularx}{\columnwidth}{l*{6}{>{\centering\arraybackslash}X}} 
    \toprule    
    Time & Fedavg & Krum & Trimmed
    \_mean & Fang & GANsweep & GANcrop \\    
    \midrule    
    s & 258 & 303 & 302 & 533 & 451 & 477 \\    
    \bottomrule    
  \end{tabularx}    
\end{table}

\section{Conclusions}

This paper addresses the problem of backdoor attacks in federated learning and proposes a defense method based on contrastive learning and Generative Adversarial Networks, GANcrop. Model comparison and sensitivity analysis of parameters effectively distinguish malicious and benign models. Then, utilizing GAN technology to recover and mitigate backdoor triggers in models, significantly reduces the success rate of backdoor attacks. Experiments have proven that compared to existing methods, GANcrop not only enhances the defense against backdoor attacks but also mitigates the risk of backdoor attacks in non-IID data scenarios while maintaining the accuracy of the main task of federated learning models. In future work, efforts will be directed towards further optimizing the mitigation strategy of GANcrop, aiming to enhance the post-mitigation model accuracy and thereby augmenting the method's practical applicability.

\section{Acknowledgments}
The research was supported in part by the Guangxi Science and Technology Major Project (No.AA22068070), the National Natural Science Foundation of China (Nos.62166004,U21A20474), the Key Lab of Education Blockchain and Intelligent Technology, the Center for Applied Mathematics of Guangxi, the Guangxi "Bagui Scholar" Teams for Innovation and Research Project, the Guangxi Talent Highland Project of Big Data Intelligence and Application, the Guangxi Collaborative Center of Multisource Information Integration and Intelligent Processing.




\bibliographystyle{ACM-Reference-Format}
\bibliography{sample-manuscript}


\begin{thebibliography}{19}


\ifx \showCODEN    \undefined \def \showCODEN     #1{\unskip}     \fi
\ifx \showDOI      \undefined \def \showDOI       #1{#1}\fi
\ifx \showISBNx    \undefined \def \showISBNx     #1{\unskip}     \fi
\ifx \showISBNxiii \undefined \def \showISBNxiii  #1{\unskip}     \fi
\ifx \showISSN     \undefined \def \showISSN      #1{\unskip}     \fi
\ifx \showLCCN     \undefined \def \showLCCN      #1{\unskip}     \fi
\ifx \shownote     \undefined \def \shownote      #1{#1}          \fi
\ifx \showarticletitle \undefined \def \showarticletitle #1{#1}   \fi
\ifx \showURL      \undefined \def \showURL       {\relax}        \fi
\providecommand\bibfield[2]{#2}
\providecommand\bibinfo[2]{#2}
\providecommand\natexlab[1]{#1}
\providecommand\showeprint[2][]{arXiv:#2}

\bibitem[Bagdasaryan et~al\mbox{.}(2020)]%
        {4}
\bibfield{author}{\bibinfo{person}{Eugene Bagdasaryan}, \bibinfo{person}{Andreas Veit}, \bibinfo{person}{Yiqing Hua}, \bibinfo{person}{Deborah Estrin}, {and} \bibinfo{person}{Vitaly Shmatikov}.} \bibinfo{year}{2020}\natexlab{}.
\newblock \showarticletitle{How To Backdoor Federated Learning}. In \bibinfo{booktitle}{\emph{Proceedings of the Twenty Third International Conference on Artificial Intelligence and Statistics}} \emph{(\bibinfo{series}{Proceedings of Machine Learning Research}, Vol.~\bibinfo{volume}{108})}, \bibfield{editor}{\bibinfo{person}{Silvia Chiappa} {and} \bibinfo{person}{Roberto Calandra}} (Eds.). \bibinfo{publisher}{PMLR}, \bibinfo{pages}{2938--2948}.
\newblock
\urldef\tempurl%
\url{https://proceedings.mlr.press/v108/bagdasaryan20a.html}
\showURL{%
\tempurl}


\bibitem[Blanchard et~al\mbox{.}(2017)]%
        {16}
\bibfield{author}{\bibinfo{person}{Peva Blanchard}, \bibinfo{person}{El~Mahdi El~Mhamdi}, \bibinfo{person}{Rachid Guerraoui}, {and} \bibinfo{person}{Julien Stainer}.} \bibinfo{year}{2017}\natexlab{}.
\newblock \showarticletitle{Machine learning with adversaries: byzantine tolerant gradient descent}. In \bibinfo{booktitle}{\emph{Proceedings of the 31st International Conference on Neural Information Processing Systems}} (Long Beach, California, USA) \emph{(\bibinfo{series}{NIPS'17})}. \bibinfo{publisher}{Curran Associates Inc.}, \bibinfo{pages}{118–128}.
\newblock
\showISBNx{9781510860964}


\bibitem[Chen et~al\mbox{.}(2020)]%
        {12}
\bibfield{author}{\bibinfo{person}{Ting Chen}, \bibinfo{person}{Simon Kornblith}, \bibinfo{person}{Mohammad Norouzi}, {and} \bibinfo{person}{Geoffrey Hinton}.} \bibinfo{year}{2020}\natexlab{}.
\newblock \bibinfo{title}{A Simple Framework for Contrastive Learning of Visual Representations}.
\newblock
\newblock
\showeprint[arxiv]{2002.05709}~[cs.LG]


\bibitem[Fung et~al\mbox{.}(2020)]%
        {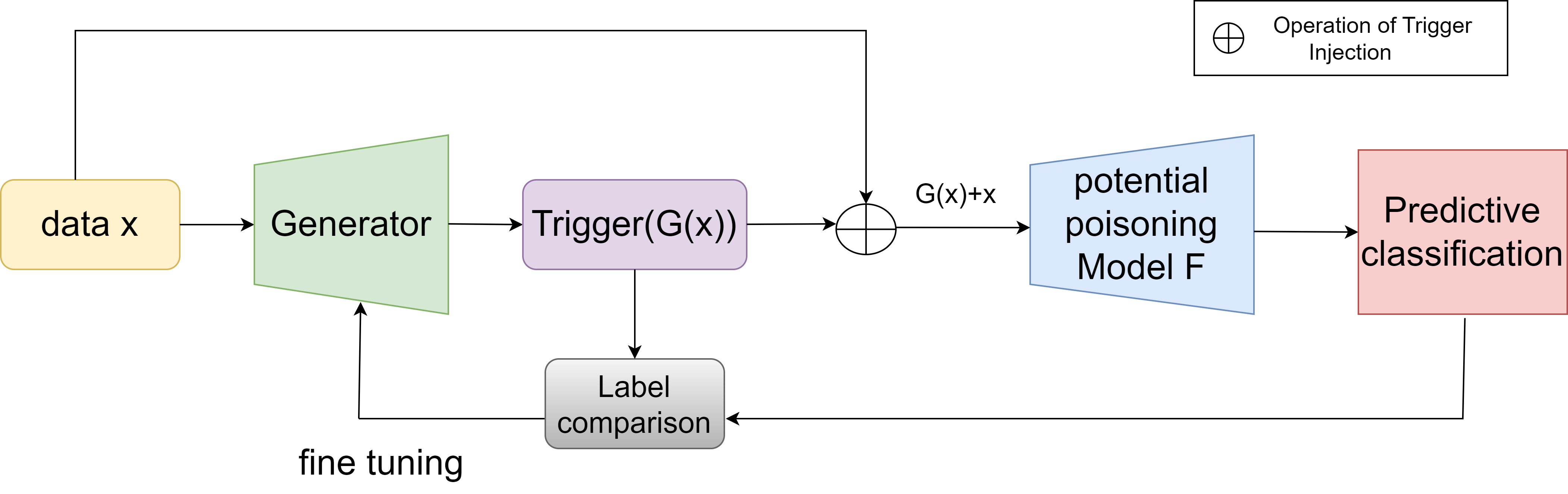}
\bibfield{author}{\bibinfo{person}{Clement Fung}, \bibinfo{person}{Chris J.~M. Yoon}, {and} \bibinfo{person}{Ivan Beschastnikh}.} \bibinfo{year}{2020}\natexlab{}.
\newblock \showarticletitle{The Limitations of Federated Learning in Sybil Settings}. In \bibinfo{booktitle}{\emph{International Symposium on Recent Advances in Intrusion Detection}}.
\newblock
\urldef\tempurl%
\url{https://api.semanticscholar.org/CorpusID:221542915}
\showURL{%
\tempurl}


\bibitem[Gao et~al\mbox{.}(2021)]%
        {7}
\bibfield{author}{\bibinfo{person}{T Gao}, \bibinfo{person}{X Yao}, {and} \bibinfo{person}{D Chen}.} \bibinfo{year}{2021}\natexlab{}.
\newblock \showarticletitle{Simcse: Simple contrastive learning of sentence embeddings}.
\newblock \bibinfo{journal}{\emph{arXiv preprint arXiv:2104.08821}} (\bibinfo{year}{2021}).
\newblock
\urldef\tempurl%
\url{https://arxiv.org/abs/2104.08821}
\showURL{%
\tempurl}


\bibitem[Goodfellow et~al\mbox{.}(2020)]%
        {13}
\bibfield{author}{\bibinfo{person}{Ian Goodfellow}, \bibinfo{person}{Jean Pouget-Abadie}, \bibinfo{person}{Mehdi Mirza}, \bibinfo{person}{Bing Xu}, \bibinfo{person}{David Warde-Farley}, \bibinfo{person}{Sherjil Ozair}, \bibinfo{person}{Aaron Courville}, {and} \bibinfo{person}{Yoshua Bengio}.} \bibinfo{year}{2020}\natexlab{}.
\newblock \showarticletitle{Generative adversarial networks}.
\newblock  \bibinfo{volume}{63}, \bibinfo{number}{11} (\bibinfo{date}{oct} \bibinfo{year}{2020}), \bibinfo{pages}{139–144}.
\newblock
\showISSN{0001-0782}


\bibitem[Li et~al\mbox{.}(2022)]%
        {15}
\bibfield{author}{\bibinfo{person}{Qinbin Li}, \bibinfo{person}{Yiqun Diao}, \bibinfo{person}{Quan Chen}, {and} \bibinfo{person}{Bingsheng He}.} \bibinfo{year}{2022}\natexlab{}.
\newblock \showarticletitle{Federated Learning on Non-IID Data Silos: An Experimental Study}. In \bibinfo{booktitle}{\emph{2022 IEEE 38th International Conference on Data Engineering (ICDE)}}. \bibinfo{pages}{965--978}.
\newblock
\urldef\tempurl%
\url{https://doi.org/10.1109/ICDE53745.2022.00077}
\showDOI{\tempurl}


\bibitem[Li et~al\mbox{.}(2019)]%
        {18}
\bibfield{author}{\bibinfo{person}{Suyi Li}, \bibinfo{person}{Yong Cheng}, \bibinfo{person}{Yang Liu}, \bibinfo{person}{Wei Wang}, {and} \bibinfo{person}{Tianjian Chen}.} \bibinfo{year}{2019}\natexlab{}.
\newblock \bibinfo{title}{Abnormal Client Behavior Detection in Federated Learning}.
\newblock
\newblock
\showeprint[arxiv]{1910.09933}~[cs.LG]


\bibitem[Li et~al\mbox{.}(2021)]%
        {19}
\bibfield{author}{\bibinfo{person}{Yiming Li}, \bibinfo{person}{Tongqing Zhai}, \bibinfo{person}{Baoyuan Wu}, \bibinfo{person}{Yong Jiang}, \bibinfo{person}{Zhifeng Li}, {and} \bibinfo{person}{Shutao Xia}.} \bibinfo{year}{2021}\natexlab{}.
\newblock \bibinfo{title}{Rethinking the Trigger of Backdoor Attack}.
\newblock
\newblock
\showeprint[arxiv]{2004.04692}~[cs.CR]


\bibitem[Liu et~al\mbox{.}(2018)]%
        {10}
\bibfield{author}{\bibinfo{person}{K Liu}, \bibinfo{person}{B Dolan-Gavitt}, {and} \bibinfo{person}{S Garg}.} \bibinfo{year}{2018}\natexlab{}.
\newblock \showarticletitle{Fine-pruning: Defending against backdooring attacks on deep neural networks}. In \bibinfo{booktitle}{\emph{International symposium on research in attacks, intrusions, and defenses}}. \bibinfo{publisher}{Springer International Publishing}, \bibinfo{pages}{273--294}.
\newblock
\urldef\tempurl%
\url{https://arxiv.org/abs/1805.12185}
\showURL{%
\tempurl}


\bibitem[McMahan et~al\mbox{.}(2017)]%
        {1}
\bibfield{author}{\bibinfo{person}{Brendan McMahan}, \bibinfo{person}{Eider Moore}, \bibinfo{person}{Daniel Ramage}, \bibinfo{person}{Seth Hampson}, {and} \bibinfo{person}{Blaise Aguera~y Arcas}.} \bibinfo{year}{2017}\natexlab{}.
\newblock \showarticletitle{{Communication-Efficient Learning of Deep Networks from Decentralized Data}}. In \bibinfo{booktitle}{\emph{Proceedings of the 20th International Conference on Artificial Intelligence and Statistics}} \emph{(\bibinfo{series}{Proceedings of Machine Learning Research}, Vol.~\bibinfo{volume}{54})}. \bibinfo{publisher}{PMLR}, \bibinfo{pages}{1273--1282}.
\newblock
\urldef\tempurl%
\url{https://proceedings.mlr.press/v54/mcmahan17a.html}
\showURL{%
\tempurl}


\bibitem[Melis et~al\mbox{.}(2018)]%
        {17}
\bibfield{author}{\bibinfo{person}{Luca Melis}, \bibinfo{person}{Congzheng Song}, \bibinfo{person}{Emiliano~De Cristofaro}, {and} \bibinfo{person}{Vitaly Shmatikov}.} \bibinfo{year}{2018}\natexlab{}.
\newblock \bibinfo{title}{Exploiting Unintended Feature Leakage in Collaborative Learning}.
\newblock
\newblock
\showeprint[arxiv]{1805.04049}~[cs.CR]


\bibitem[Nguyen et~al\mbox{.}(2024)]%
        {14}
\bibfield{author}{\bibinfo{person}{Thuy~Dung Nguyen}, \bibinfo{person}{Tuan Nguyen}, \bibinfo{person}{Phi~Le Nguyen}, \bibinfo{person}{Hieu~H. Pham}, \bibinfo{person}{Khoa~D. Doan}, {and} \bibinfo{person}{Kok-Seng Wong}.} \bibinfo{year}{2024}\natexlab{}.
\newblock \showarticletitle{Backdoor attacks and defenses in federated learning: Survey, challenges and future research directions}.
\newblock \bibinfo{journal}{\emph{Engineering Applications of Artificial Intelligence}}  \bibinfo{volume}{127} (\bibinfo{year}{2024}), \bibinfo{pages}{107166}.
\newblock
\showISSN{0952-1976}
\urldef\tempurl%
\url{https://doi.org/10.1016/j.engappai.2023.107166}
\showDOI{\tempurl}


\bibitem[Radford et~al\mbox{.}(2021)]%
        {9}
\bibfield{author}{\bibinfo{person}{Alec Radford}, \bibinfo{person}{Jong~Wook Kim}, \bibinfo{person}{Chris Hallacy}, \bibinfo{person}{Aditya Ramesh}, \bibinfo{person}{Gabriel Goh}, \bibinfo{person}{Sandhini Agarwal}, \bibinfo{person}{Girish Sastry}, \bibinfo{person}{Amanda Askell}, \bibinfo{person}{Pamela Mishkin}, \bibinfo{person}{Jack Clark}, \bibinfo{person}{Gretchen Krueger}, {and} \bibinfo{person}{Ilya Sutskever}.} \bibinfo{year}{2021}\natexlab{}.
\newblock \bibinfo{title}{Learning Transferable Visual Models From Natural Language Supervision}.
\newblock
\newblock
\showeprint[arxiv]{2103.00020}~[cs.CV]


\bibitem[Sun et~al\mbox{.}(2019)]%
        {11}
\bibfield{author}{\bibinfo{person}{Ziteng Sun}, \bibinfo{person}{Peter Kairouz}, \bibinfo{person}{Ananda~Theertha Suresh}, {and} \bibinfo{person}{H.~Brendan McMahan}.} \bibinfo{year}{2019}\natexlab{}.
\newblock \bibinfo{title}{Can You Really Backdoor Federated Learning?}
\newblock
\newblock
\showeprint[arxiv]{1911.07963}~[cs.LG]


\bibitem[Wang et~al\mbox{.}(2019)]%
        {5}
\bibfield{author}{\bibinfo{person}{Bolun Wang}, \bibinfo{person}{Yuanshun Yao}, \bibinfo{person}{Shawn Shan}, \bibinfo{person}{Huiying Li}, \bibinfo{person}{Bimal Viswanath}, \bibinfo{person}{Haitao Zheng}, {and} \bibinfo{person}{Ben~Y. Zhao}.} \bibinfo{year}{2019}\natexlab{}.
\newblock \showarticletitle{Neural Cleanse: Identifying and Mitigating Backdoor Attacks in Neural Networks}. In \bibinfo{booktitle}{\emph{2019 IEEE Symposium on Security and Privacy (SP)}}. \bibinfo{pages}{707--723}.
\newblock
\urldef\tempurl%
\url{https://doi.org/10.1109/SP.2019.00031}
\showDOI{\tempurl}


\bibitem[Xie et~al\mbox{.}(2020)]%
        {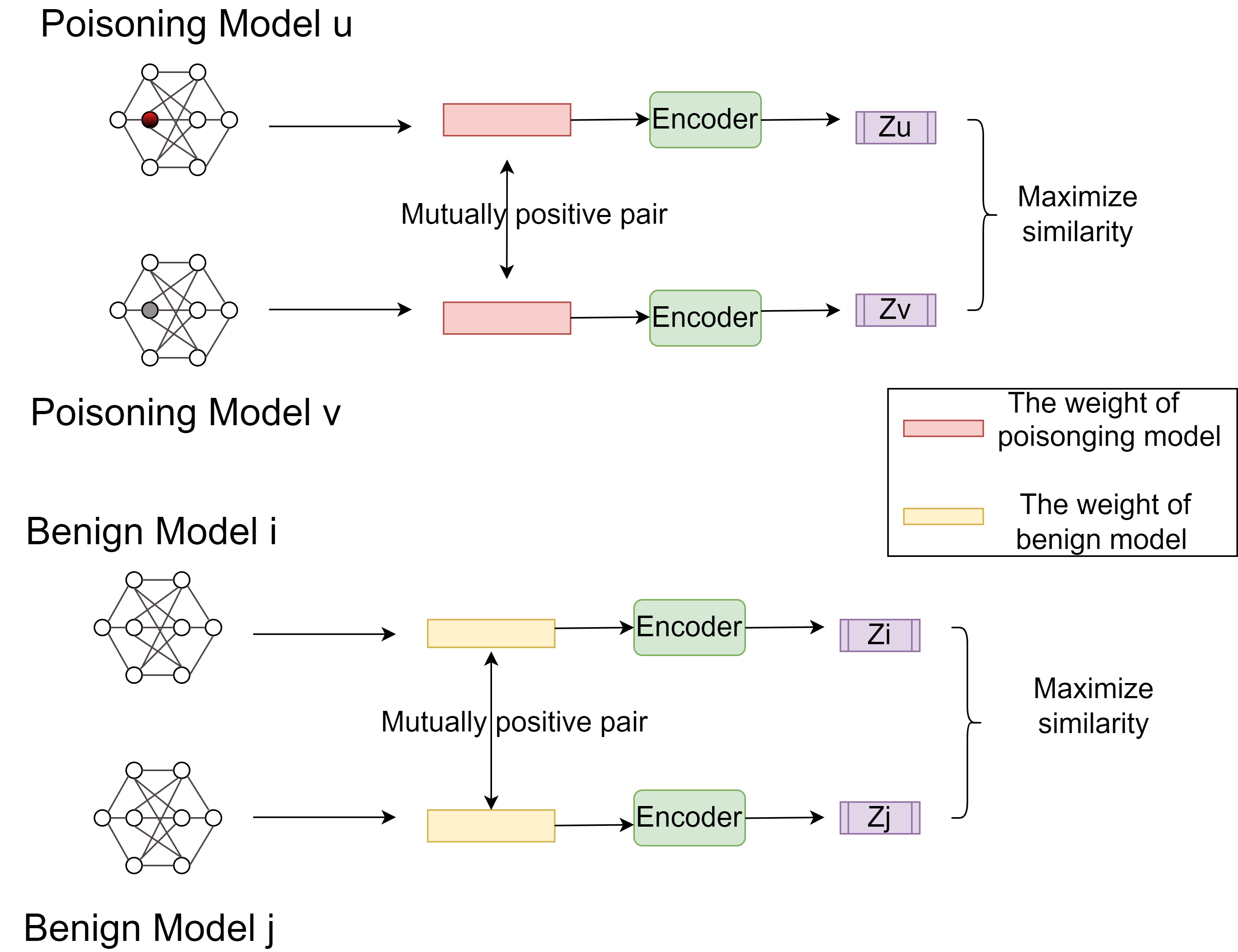}
\bibfield{author}{\bibinfo{person}{Chulin Xie}, \bibinfo{person}{Keli Huang}, \bibinfo{person}{Pin-Yu Chen}, {and} \bibinfo{person}{Bo Li}.} \bibinfo{year}{2020}\natexlab{}.
\newblock \showarticletitle{DBA: Distributed Backdoor Attacks against Federated Learning}. In \bibinfo{booktitle}{\emph{International Conference on Learning Representations}}.
\newblock
\urldef\tempurl%
\url{https://openreview.net/forum?id=rkgyS0VFvr}
\showURL{%
\tempurl}


\bibitem[Ye et~al\mbox{.}(2023)]%
        {8}
\bibfield{author}{\bibinfo{person}{Mang Ye}, \bibinfo{person}{Xiuwen Fang}, \bibinfo{person}{Bo Du}, \bibinfo{person}{Pong~C. Yuen}, {and} \bibinfo{person}{Dacheng Tao}.} \bibinfo{year}{2023}\natexlab{}.
\newblock \bibinfo{title}{Heterogeneous Federated Learning: State-of-the-art and Research Challenges}.
\newblock
\newblock
\showeprint[arxiv]{2307.10616}~[cs.LG]


\bibitem[Zhu et~al\mbox{.}(2020)]%
        {6}
\bibfield{author}{\bibinfo{person}{Liuwan Zhu}, \bibinfo{person}{Rui Ning}, \bibinfo{person}{Cong Wang}, \bibinfo{person}{Chunsheng Xin}, {and} \bibinfo{person}{Hongyi Wu}.} \bibinfo{year}{2020}\natexlab{}.
\newblock \showarticletitle{GangSweep: Sweep out Neural Backdoors by GAN}. In \bibinfo{booktitle}{\emph{Proceedings of the 28th ACM International Conference on Multimedia}} \emph{(\bibinfo{series}{MM '20})}. \bibinfo{pages}{3173–3181}.
\newblock
\showISBNx{9781450379885}
\urldef\tempurl%
\url{https://doi.org/10.1145/3394171.3413546}
\showURL{%
\tempurl}


\end{thebibliography}

\end{document}